\documentclass[useAMS,usenatbib,usegraphicx]{mn2e}
\usepackage{psfig}   
\usepackage{graphicx}
\usepackage{subfigure}
\usepackage{amssymb}
\usepackage{amsmath}
\usepackage{txfonts}
\usepackage{longtable}

\title[Dynamics versus structure]{Dynamics versus structure: breaking the density degeneracy in star formation}

\author[R.~J.~Parker]{
  Richard J.~Parker\thanks{E-mail: R.J.Parker@ljmu.ac.uk}
  \vspace*{0.1cm}\\
Astrophysics Research Institute, Liverpool John Moores University, 146 Brownlow Hill, Liverpool, L3 5RF, UK}

\begin{document}

\date{}
                             
\pagerange{\pageref{firstpage}--\pageref{lastpage}} \pubyear{2014}

\maketitle

\label{firstpage}

\begin{abstract}
The initial density of individual star-forming regions (and by extension the birth environment of planetary systems) is difficult to constrain due to the ``density degeneracy problem'': an initially dense region expands faster than a more quiescent region due to two-body relaxation and so two regions with the same observed present-day density may have had very different initial densities. We constrain the initial densities of seven nearby star-forming regions by folding in information on their spatial structure from the $\mathcal{Q}$-parameter and comparing the structure and present-day density to the results of $N$-body simulations. This in turn places strong constraints on the possible effects of dynamical interactions and radiation fields from massive stars on multiple systems and protoplanetary discs.

We apply our method to constrain the initial binary population in each of these seven regions and show that the populations in only three -- the Orion Nebula Cluster, $\rho$~Oph and Corona Australis -- are consistent with having evolved from the Kroupa universal initial period distribution and a binary fraction of unity.
\end{abstract}


\section{Introduction}

Characterising the formation environment of stars is one of the outstanding challenges in astrophysics. If stars are predominately born in dense `clustered' environments \citep[e.g.][]{Lada03,Lada10} then dynamical interactions and the radiation fields from massive stars  may significantly affect the formation and evolution of planetary systems \citep[e.g.][]{Armitage00,Bonnell01,Scally01,Adams06,Olczak08,Parker12a,Rosotti14} and the properties of binary and multiple systems \citep[e.g.][]{Kroupa95a,Kroupa99,Marks12,Parker12b}.

On the other hand, if most stars are born in relative isolation \citep*[e.g.][]{Shu87}, or rather in low-density environments where dynamical interactions are insignificant \citep[e.g.][]{Bressert10}, then planetary and binary systems may form with little or no external perturbation. Either scenario has important implications for understanding the origin of stars in the Galactic field, and for placing our Solar System in the context of exoplanetary systems \citep[e.g.][and references therein]{Adams11,Alexander13,Davies13}.

Ideally, we would like to compare the properties of observed star-forming regions and young clusters to simulations to gauge the effects of the star-forming environment on binary systems and fledgling planetary systems. Binary systems are particularly useful because their properties in the Galactic field are well-constrained \citep{Raghavan10,Duchene13b}. In principle one can compare binary populations between star-forming regions and the Galactic field, in order to determine the type of star-forming region that produces the most `field-like' binaries \citep[and hence is the dominant star-forming event that produces the Galactic field;][]{Goodwin10}.

Unfortunately, this problem is severely complicated by uncertainty in determining the maximum density attained by star-forming regions. Observations of the present-day density in star-forming regions provide very few constraints on the initial density \citep[e.g.][]{King12a,Moeckel12,Gieles12,Parker12d}. The reason is that an initially dense region expands very quickly due to two-body relaxation, whereas a less dense region expands more slowly. Therefore, at a given age two regions with the same present-day density may have had very different densities in the past. This is the so-called ``density degeneracy problem'' -- not enough information is available to rule out much more dense initial conditions \citep[e.g.][]{Marks12,Marks14}. 

In this paper, we attempt to address this issue by folding in extra information on the structure of star forming regions \citep{Cartwright04}, and (where available) the relative density around massive stars with respect to the median stellar density in the region \citep{Maschberger11}. We compare observational data for seven nearby star-forming regions to the results of $N$-body simulations where we vary the initial density to determine the most likely initial conditions of each region. As an example of the method, we then use these constraints to rule out the universal binary population hypothesis from \citet{Kroupa95a,Kroupa95b}. We describe our simulation set-up in Section~\ref{method}, we present our results in Section~\ref{results} and we conclude in Section~\ref{conclude}.

\section{Method}
\label{method}

\subsection{Star-forming region set-up}

Both observations \citep[e.g.][]{Cartwright04,Sanchez09,Gouliermis14} and simulations \citep{Schmeja06,Girichidis12,Dale13} of star-forming regions indicate that stars form with a hierarchical, or self-similar spatial distribution (i.e.\,\,they are substructured). It is almost impossible to create substructure through dynamical interactions; rather it is usually completely erased over a few crossing times \citep{Parker14b}. Therefore, in order to reproduce the substructure observed in many of the regions of interest here, we must start the simulations with substructure.

We set up substructured star forming regions using fractal distributions, following the method of \citet{Goodwin04a}. This method is described in detail in that paper, and in \citet{Allison10} and \citet{Parker14b}. Briefly, the fractal is built by creating a cube containing `parents', which spawn a number of `children' depending on the desired fractal dimension. The amount of substructure is then set by the number of children that are allowed to mature. The lower the fractal dimension, the fewer children are allowed to mature and the cube has more substructure. Fractal dimensions in the range $D = 1.6$ (highly substructured) to $D = 3.0$ (uniform distribution) are allowed. Finally, outlying particles are removed so that the cube from which the fractal was created becomes a sphere; however, the distribution is only truly spherical if $D = 3.0$. 

All of our simulated star-forming regions have a fractal dimension $D = 1.6$; the Taurus association \citep{Cartwright04} and Corona Australis \citep[CrA,][]{Neuhauser08} both have fractal dimensions consistent with this value and because dynamical interactions cannot make a region more substructured, we adopt this value. However, we note that hydrodynamical simulations of star formation can produce less substructured regions (higher $D$ values) and as we shall see, some observed regions must also have higher primordial fractal dimensions. 

The velocities of stars in the fractals are also correlated on local scales, in accordance with observations \citep{Larson82,Andre10}. The children in our fractals inherit their parents' velocity, plus a small amount of noise which successively decreases further down the fractal tree. This means that two nearby stars have very similar velocities, whereas two stars which are distant can have very different velocities. Again, this is an effort to mimic the observations of star formation, which indicate that stars in filaments have very low velocity dispersions \citep{Andre10}.  

In order to erase primordial substructure and to process primordial binary systems as efficiently as possible, we scale the velocities of the whole fractal to be subvirial ($\alpha_{\rm vir} = 0.3$, where the virial ratio $\alpha_{\rm vir} = T/|\Omega|$; $T$ and $\Omega$ are the total kinetic energy and total potential energy of the stars, respectively). 

We set up our star-forming regions with three different densities. In two sets of simulations, the regions have a radius of 1\,pc and contain either 1500 stars (which we will refer to as ``high density'' -- $\tilde{\rho} \sim10^4$\,M$_\odot$\,pc$^{-3}$) or 150 stars (``medium density'' -- $\tilde{\rho} \sim10^2$\,M$_\odot$\,pc$^{-3}$). In a third set of simulations, the regions contain 300 stars and have a radius of 5\,pc (``low density'' -- $\tilde{\rho} \sim10$\,M$_\odot$\,pc$^{-3}$).

\subsection{Binary population}

All of our regions have an initial binary fraction of unity, i.e.\,\,everything forms in a binary. When creating the binary populations we adopt the same initial conditions as in \citet{Marks14}. The primary masses are drawn from a \citet{Kroupa02} IMF of the form 
\begin{equation}
 \frac{dN}{dM}   \propto  \left\{ \begin{array}{ll} M^{-1.3} \hspace{0.4cm} m_0
  < M/{\rm M_\odot} \leq m_1   \,, \\ M^{-2.3} \hspace{0.4cm} m_1 <
  M/{\rm M_\odot} \leq m_2   \,,
\end{array} \right.
\end{equation}
where $m_0$ = 0.1\,M$_\odot$, $m_1$ = 0.5\,M$_\odot$, and  $m_2$ = 50\,M$_\odot$.
clusters. There are no brown dwarfs in the simulations. Secondary masses are also drawn at random from the IMF; note this is inconsistent with recent observations \citep{Metchev09,Reggiani13} which show a universal flat companion mass ration distribution. However, subsequent pre-main sequence eigenevolution (see below) alters the mass ratios of close binaries so that the CMRD approaches a flat distribution. 

Binary periods are drawn from the \citet{Kroupa95b} period distribution \citep[see also][]{Kroupa11,Marks14} of the form 
\begin{equation}
f\left({\rm log_{10}}P\right) = \eta\frac{{\rm log_{10}}P - {\rm log_{10}}P_{\rm min}}{\delta + \left({\rm log_{10}}P - {\rm log_{10}}P_{\rm min}\right)^2
},
\label{period}
\end{equation}
where ${\rm log_{10}}P_{\rm min}$ is the logarithm of the minimum period in days and ${\rm log_{10}}P_{\rm min} = 1$. $\eta = 2.5$ and $\delta = 45$ are 
the numerical constants adopted by \citet{Kroupa95b}. This period distribution was derived from a process of ``reverse engineering'' $N$-body simulations  \citep{Kroupa95a,Kroupa95b,Kroupa95c}; regions with low densities do not break up many binaries and hence would have an excess of wide systems (100 -- 10$^4$\,au) compared to the Galactic field, as observed in Taurus \citep{Leinert93,Kohler98}, whereas more dense regions would destroy more wider binaries and the resultant separation distribution is more ``field-like'' \citep{Duquennoy91,Fischer92,Raghavan10}.

Eccentricities are drawn from a thermal distribution \citep{Heggie75} of the form
\begin{equation}
f(e) = 2e.
\end{equation} 
We note that the eccentricity distribution of binaries in the field is more consistent with a flat distribution \citep{Raghavan10,Duchene13b}; however, as with the mass ratios, eigenevolution alters the distribution for close systems.

Finally, we apply the \citet{Kroupa95b} `eigenevolution' algorithm, which accounts for tidal circularisation effects in close binaries \citep{Mathieu94}, and for early angular momentum transfer between the circumprimary disk and the secondary star.

We then place the binary systems at the centre of mass of each position in the fractal and we evolve the star-forming regions for 10\,Myr using the \texttt{kira} integrator in the \texttt{Starlab} package \citep{Zwart99,Zwart01}. We do not include stellar evolution in the simulations. 

\section{Results}
\label{results}

\begin{figure}
\begin{center}
\rotatebox{270}{\includegraphics[scale=0.4]{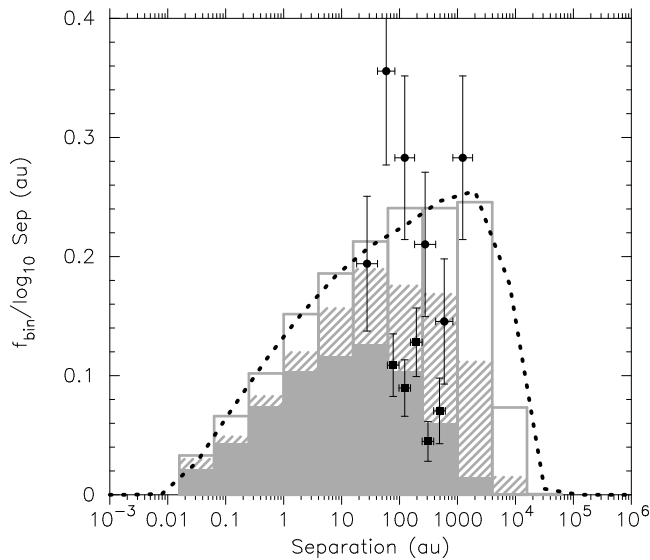}}
\end{center}
\caption[bf]{Evolution of the separation distribution normalised to the binary fraction  as a function of initial density. The primordial distribution \citep{Kroupa95b} is shown by the dotted line and the distributions after 1\,Myr are shown by the open (low density -- $\tilde{\rho} \sim10$\,M$_\odot$\,pc$^{-3}$), hashed (medium density -- $\tilde{\rho} \sim10^2$\,M$_\odot$\,pc$^{-3}$) and solid (high density -- $\tilde{\rho} \sim10^4$\,M$_\odot$\,pc$^{-3}$) histograms. The observed distributions for Taurus \citep[the circles,][]{Kohler98} and the Orion Nebula Cluster \citep[the squares,][]{Reipurth07} are also shown.}
\label{separations}
\end{figure}

We first demonstrate the density degeneracy and its effect on the binary properties in star-forming regions. In order to invoke a universal model of star formation and to reconcile differences between the binary populations of Taurus \citep{Leinert93} and the Galactic field \citep{Duquennoy91}, \citet{Kroupa95a,Kroupa95b} postulated a universal initial binary population where all stars form in binaries, with an excess of systems with wide ($10^2 - 10^4$\,au) semimajor axes with respect to the field population. In Fig.~\ref{separations} we show the initial \citet{Kroupa95b} binary period distribution (Eqn.~\ref{period}), converted to a separation distribution, by the dotted line. Depending on the maximum density attained by the region, the binaries can suffer none, little, or much dynamical destruction and the separation distribution is altered accordingly. We show the distribution (at 1\,Myr) in the simulated low density regions by the open histogram, the distribution in the medium density regions by the hashed histogram and the distribution in the high density regions by the solid histogram.  We also show the observational data points for Taurus  \citep[consistent with little dynamical evolution of the proposed initial period distribution;][]{Kohler98} by the circles and the Orion Nebula Cluster \citep[ONC, consistent with significant dynamical evolution of the proposed initial period distribution;][]{Reipurth07}, by the squares.

\begin{table*}
\caption[bf]{A summary of the regions with which we compare our $N$-body simulations. From left to right, the columns show the region name, age, $\mathcal{Q}$--parameter, an alternative determination where applicable, $\mathcal{Q}_{\rm alt}$ (see text for details), the $\Sigma_{\rm LDR}$ ratio (if available), the references for $\mathcal{Q}$, $\mathcal{Q}_{\rm alt}$, $\Sigma_{\rm LDR}$, the observed present-day density of each region as noted by \citet{Marks12} and \citet{King12b}, $\rho_{\rm obs.}$, the postulated initial density, $\rho_{\rm post.}$,  from \citet{Marks12} and \citet{Marks14} for the binary population of that region to be consistent with the universal primordial binary properties \citep{Kroupa95a}, and the symbol used in Figs.~\ref{density}~and~\ref{Qpar}.}
\begin{center}
\begin{tabular}{|c|c|c|c|c|c|c|c|c|}
\hline 
Region & Age & $\mathcal{Q}$ & $\mathcal{Q}_{\rm alt.}$ & $\Sigma_{\rm LDR}$  & Refs. & $\rho_{\rm obs.}$ & $\rho_{\rm post.}$ & symbol \\
\hline
ONC & 1\,Myr & 0.87 & 0.94 & 3.7 & \citealp{Hillenbrand98} & 400\,M$_\odot$\,pc$^{-3}$  & 68\,000\,M$_\odot$\,pc$^{-3}$ & $\blacksquare$ \\
$\rho$~Oph & 1\,Myr & 0.85 & 0.56 & 0.58 & \citealp{Cartwright04,Parker12c} & 200\,M$_\odot$\,pc$^{-3}$ & 2300\,M$_\odot$\,pc$^{-3}$ &  $\Diamondblack$ \\  
Taurus & 1\,Myr & 0.48 & -- & 0.28 & \citealp{Cartwright04,Parker11b} & 8\,M$_\odot$\,pc$^{-3}$  & 350\,M$_\odot$\,pc$^{-3}$ & $\bullet$ \\
IC\,348 & 3\,Myr & 0.92 & -- & -- & \citealp{Cartwright04} & 180\,M$_\odot$\,pc$^{-3}$ & 9400\,M$_\odot$\,pc$^{-3}$ & $\lozenge$ \\
Cham~I & 3\,Myr & 0.66 & 0.71 & -- & \citealp{Cartwright04,Cartwright09a} & 1\,M$_\odot$\,pc$^{-3}$ & 1600\,M$_\odot$\,pc$^{-3}$ & {\large{$\star$}}\\
CrA & 1\,Myr & 0.38 & 0.32 & -- & \citealp{Neuhauser08} & 30\,M$_\odot$\,pc$^{-3}$ & 190\,M$_\odot$\,pc$^{-3}$ & $\blacktriangle$ \\
Upper Sco & 5\,Myr$^a$ & 0.88 & 0.75 & -- & \citealp{Kraus07b} & 16\,M$_\odot$\,pc$^{-3}$ & 4200\,M$_\odot$\,pc$^{-3}$ & $\circ$ \\
\hline
\end{tabular}
\end{center}
{$^a$Note that the age of Upper Sco may be as high as 11\,Myr \citep*{Pecaut12}.}
\label{summary_regions}
\end{table*}

\subsection{Evolution of density}

\begin{figure*}
  \begin{center}
\setlength{\subfigcapskip}{10pt}
\hspace*{-0.7cm}
\subfigure[High initial density ($\tilde{\rho} \sim10^4$\,M$_\odot$\,pc$^{-3}$)]{\label{density-a}\rotatebox{270}{\includegraphics[scale=0.28]{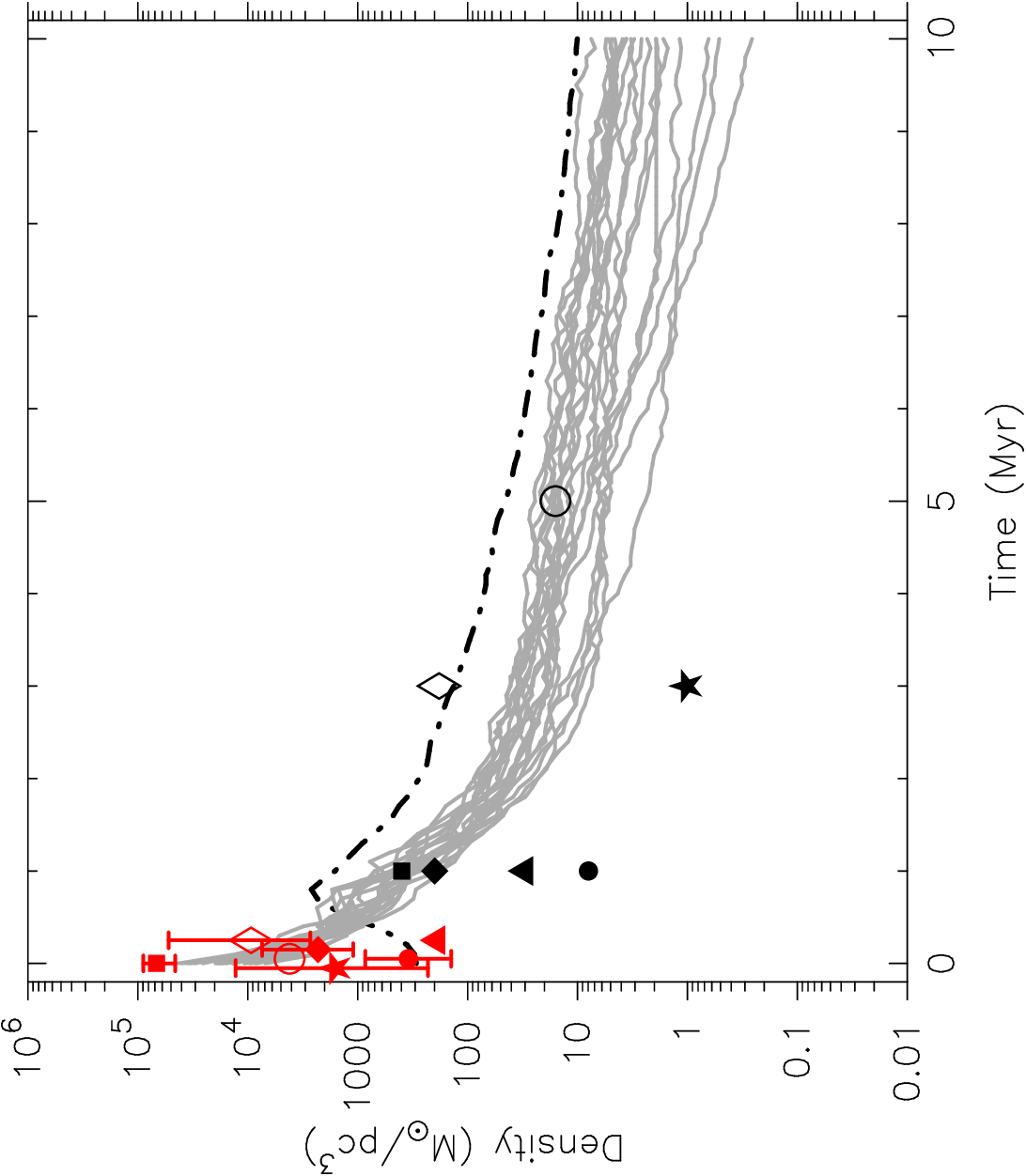}}}  
\hspace*{0.1cm}
\subfigure[Medium initial density ($\tilde{\rho} \sim10^2$\,M$_\odot$\,pc$^{-3}$)]{\label{density-b}\rotatebox{270}{\includegraphics[scale=0.28]{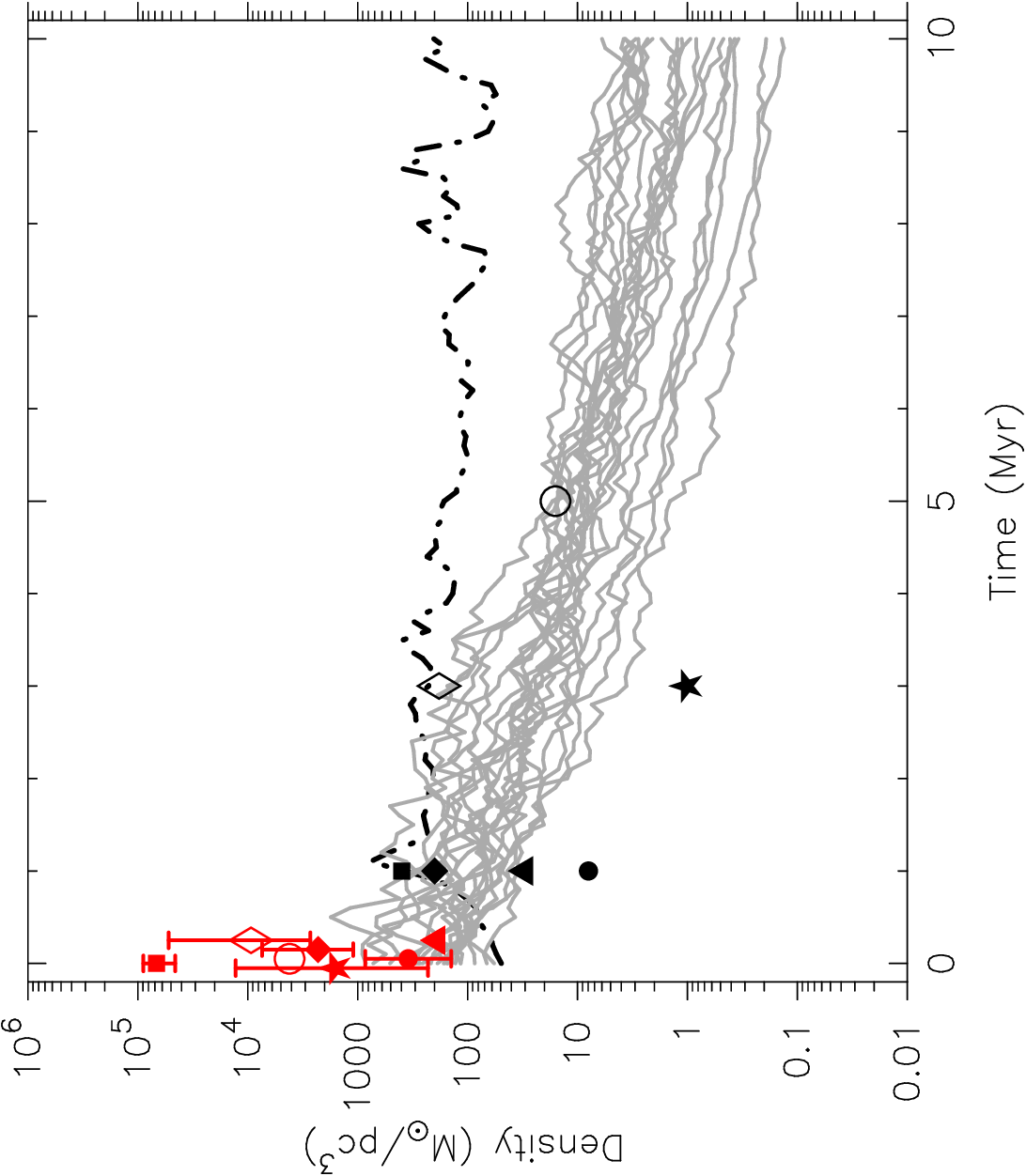}}}
\hspace*{0.1cm}
\subfigure[Low initial density ($\tilde{\rho} \sim10$\,M$_\odot$\,pc$^{-3}$)]{\label{density-c}\rotatebox{270}{\includegraphics[scale=0.28]{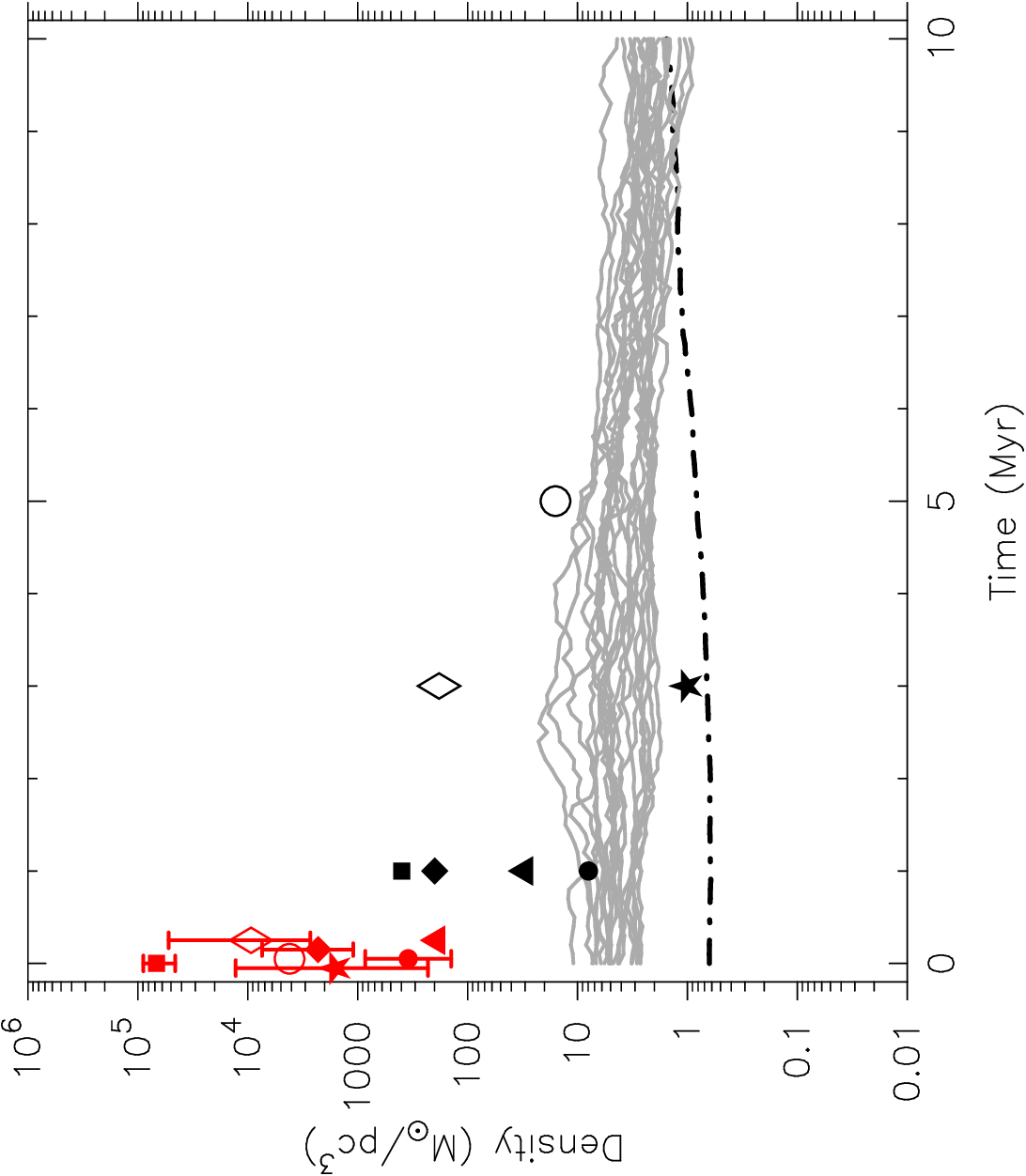}}}
\end{center}
  \caption[bf]{Evolution of the density in our simulated star-forming
    regions. In panel (a) the star-forming regions have high initial
    densities ($\tilde{\rho} \sim10^4$\,M$_\odot$\,pc$^{-3}$), in panel 
    (b) the regions have medium initial densities ($\tilde{\rho} \sim10^2$\,M$_\odot$\,pc$^{-3}$)  and in  panel (c) the regions have much lower initial densities ($\tilde{\rho} \sim10$\,M$_\odot$\,pc$^{-3}$). We show the median
    stellar volume density in each simulation by the individual grey
    (solid) lines, and the central density (within the
    half-mass radius) from twenty averaged simulations. The lefthand
    red symbols (at $t = 0$\,Myr, slightly offset from one another for clarity) are the required initial densities
    for several nearby star-forming regions if star formation is
    consistent with a
    universal initial binary population \citep{Marks12,Marks14}. The
    corresponding present-day stellar densities are shown by the black
    points at 1, 3 and 5\,Myr, depending on the age of the region. A
    key to the symbols is provided in Table~\ref{summary_regions}. 
  }
  \label{density}
\end{figure*}

\begin{figure*}
  \begin{center}
\setlength{\subfigcapskip}{10pt}
\hspace*{-0.5cm}
\subfigure[High initial density ($\tilde{\rho} \sim10^4$\,M$_\odot$\,pc$^{-3}$)]{\label{Qpar-a}\rotatebox{270}{\includegraphics[scale=0.28]{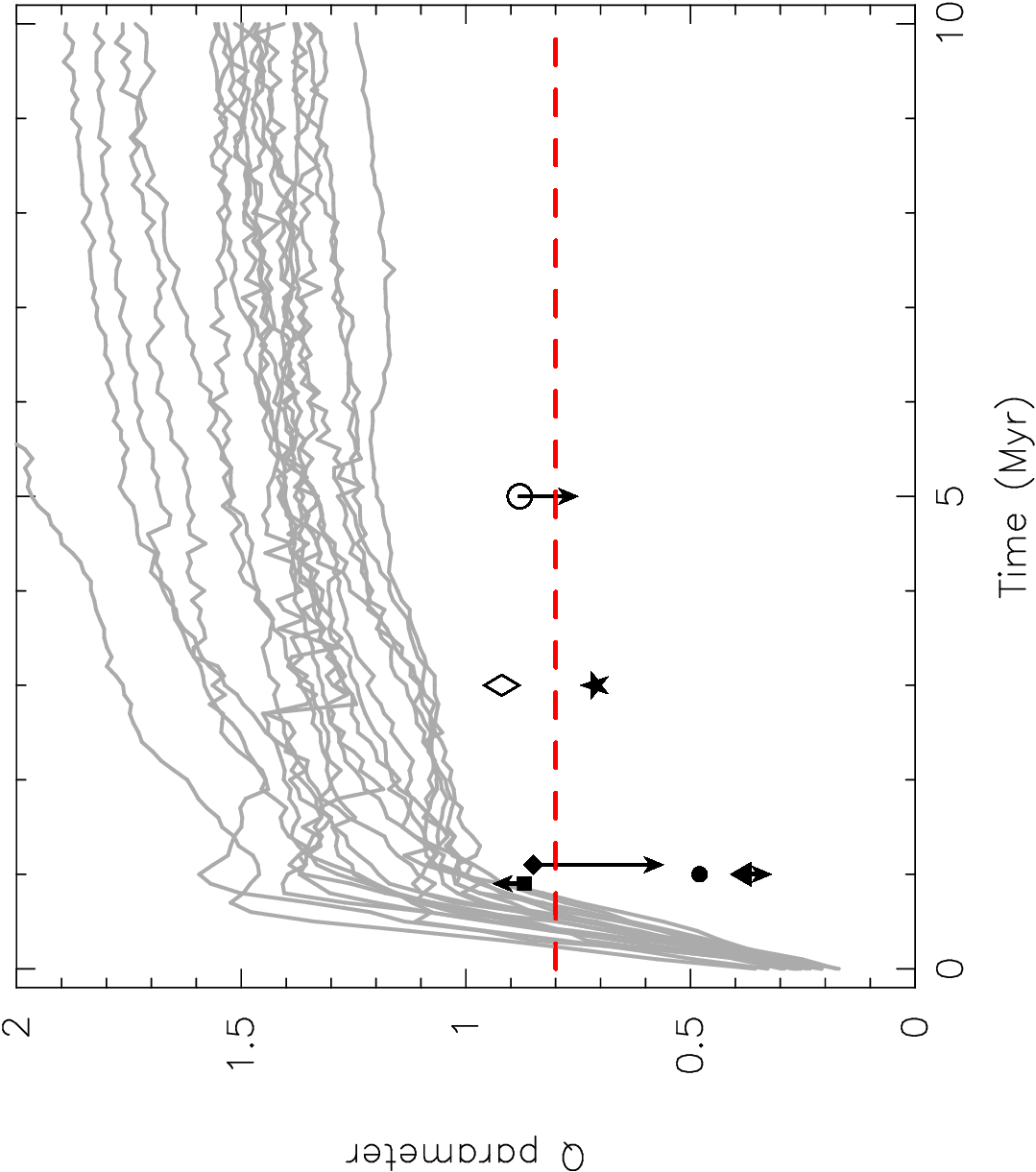}}}  
\hspace*{0.1cm}
\subfigure[Medium initial density ($\tilde{\rho} \sim10^2$\,M$_\odot$\,pc$^{-3}$)]{\label{Qpar-b}\rotatebox{270}{\includegraphics[scale=0.28]{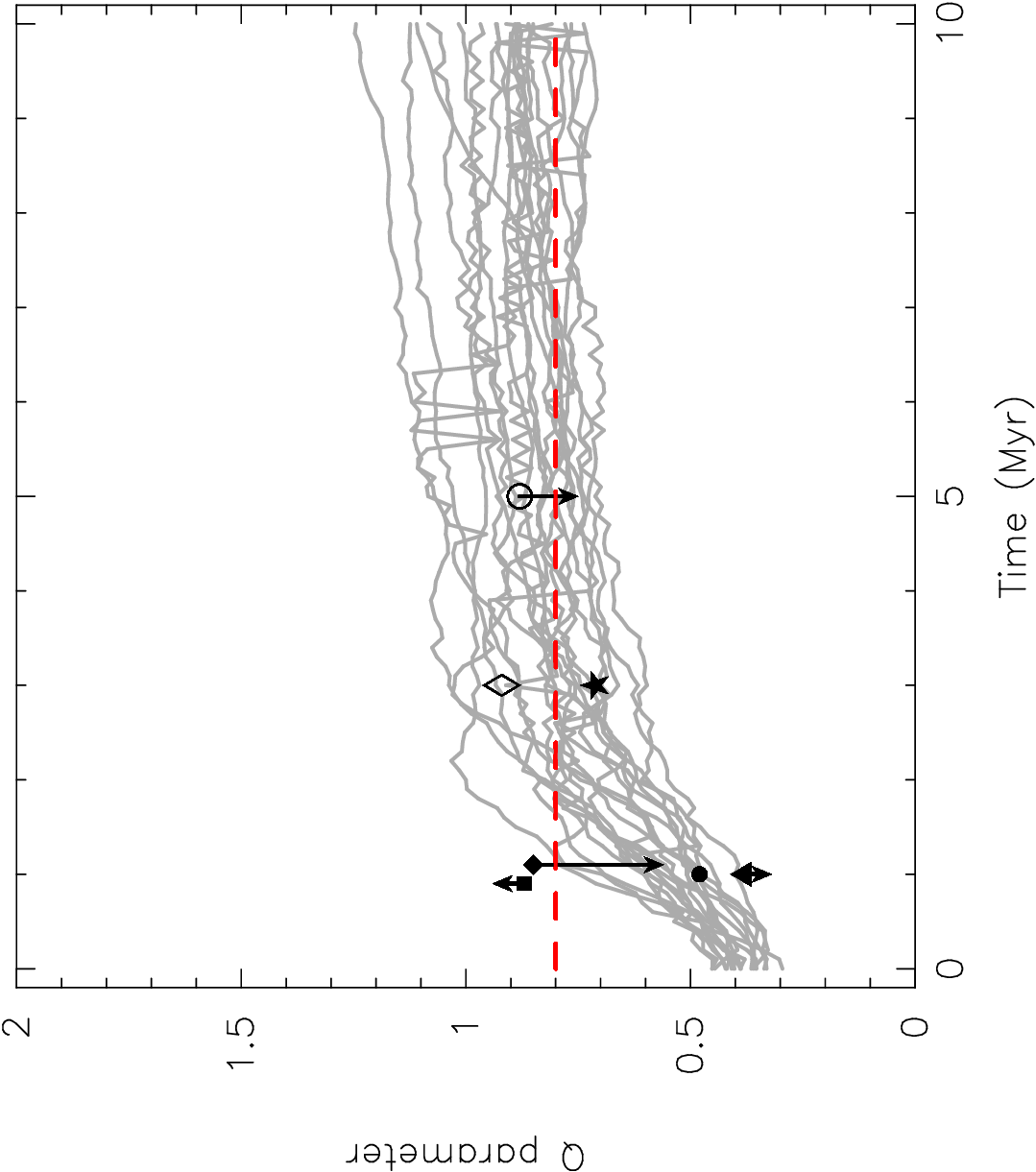}}}
\hspace*{0.1cm}
\subfigure[Low initial density ($\tilde{\rho} \sim10$\,M\,$_\odot$\,pc$^{-3}$)]{\label{Qpar-c}\rotatebox{270}{\includegraphics[scale=0.28]{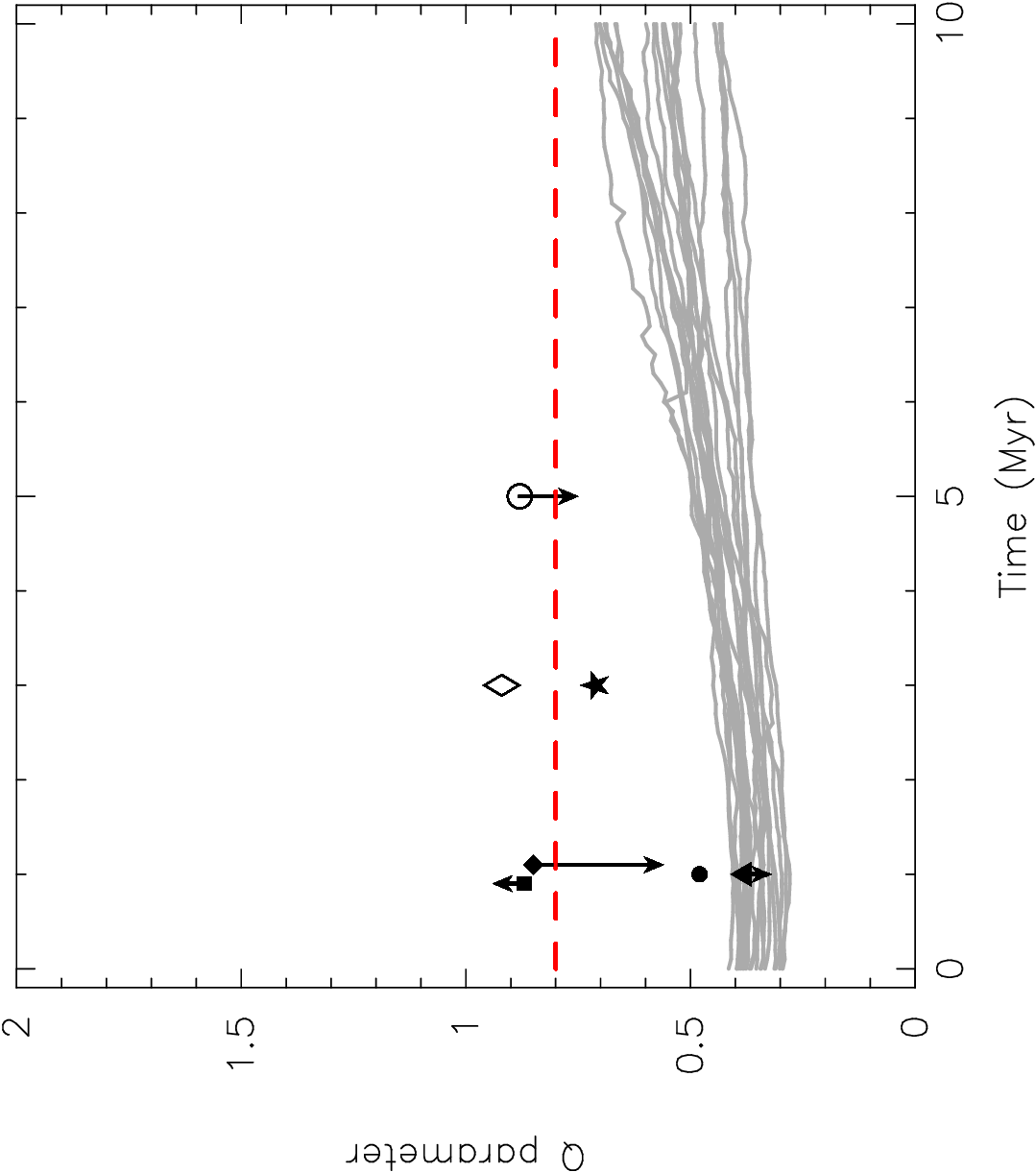}}}
\end{center}
  \caption[bf]{Evolution of structure as measured by the $\mathcal{Q}$-parameter in our simulated star-forming regions. In panel (a) the star-forming regions have high initial densities ($\tilde{\rho} \sim10^4$\,M$_\odot$\,pc$^{-3}$), in panel (b) the regions have medium initial densities ($\tilde{\rho} \sim10^2$\,M$_\odot$\,pc$^{-3}$) and in panel (c) the regions have much lower initial densities ($\tilde{\rho} \sim10$\,M$_\odot$\,pc$^{-3}$). We show the evolution of the $\mathcal{Q}$-parameter in each simulation by the individual grey (solid) lines. The boundary between substructured regions and centrally concentrated regions at $\mathcal{Q} = 0.8$ is shown by the horizontal dashed line. The $\mathcal{Q}$-parameters measured in the star-forming regions of interest are shown by the 
    points at 1, 3 and 5\,Myr, depending on the age of the
    region. Where there is an uncertainty associated with the
    measurement of $\mathcal{Q}$, we draw an arrow in the direction to
    indicate the possible deviation from the measured value. A
    key to the symbols is provided in Table~\ref{summary_regions}}
  \label{Qpar}
\end{figure*}

If a star-forming region is older, it has had more time to process its primordial binary population \citep{Marks12}. Therefore, a 3\,Myr old region can have a much lower density than a 1\,Myr old region, even though they may have had the same initial density; the difference is that two-body relaxation has caused the older region to expand more over time. We  show the evolution of density as a function of time in Fig.~\ref{density}. In panel (a) we show the evolution of our high density ($\tilde{\rho} \sim10^4$\,M$_\odot$\,pc$^{-3}$) regions, and in panel (b) we show the medium initial density ($\tilde{\rho} \sim10^2$\,M$_\odot$\,pc$^{-3}$) regions. In panel (c) we show the evolution of the low-density regions ($\tilde{\rho} \sim10$\,M$_\odot$\,pc$^{-3}$) In all panels, the median density in each of our 20 simulated regions is shown by the solid grey lines. 

The regions evolve to form a bound stellar cluster, and stars which are in the very centre of the cluster have higher densities than the region median (the grey lines). We show the evolution of the averaged central density (the volume density within the half-mass radius) by the dot-dashed lines. Conversely, stars that are ejected from the regions and become unbound have significantly lower densities. 

In Fig.~\ref{density} we also show the current density of several nearby regions of varying ages (see the final column of Table~\ref{summary_regions} for a key to the symbols). \citet{Marks12} and \citet{Marks14} argue that given a universal primordial binary population, limits can be placed on the primordial density of a region by comparing the outcome of $N$-body simulations with the currently observed visual binary population. We indicate the best-fit initial density for each region studied in \citet{Marks12} and \citet{Marks14} by the red symbols around $t = 0$\,Myr (the same symbols are used as for the present-day densities -- for example, Cham~I is shown by the {\large{$\star$}}). Outside of the error bars, \citet{Marks12} reject the possibility of that density being consistent with the processing of a common binary population with 90\,per cent confidence. 

Taking the density in isolation, Fig.~\ref{density} shows that for $\rho$~Oph and the ONC (the filled diamonds and squares, respectively) both a high-density region which evolves to far lower densities (panel a) and a medium-density region that remains static within the first Myr (panel b) are consistent with the observations. However, when their binary populations are considered, \citet{Marks12} show that under the assumption of a universal primordial binary population, the initial densities must be more than a factor of 10 different.

\subsection{Evolution of structure}

In order to break this density degeneracy, we compare the evolution of the spatial structure  in our simulations, as measured by the $\mathcal{Q}$--parameter \citep{Cartwright04,Cartwright09a,Cartwright09b}. The $\mathcal{Q}$--parameter compares the mean length of the minimum spanning tree (the shortest possible pathlength between all stars where there are no closed loops, $\bar{m}$) to the mean separation between stars, $\bar{s}$:
\begin{equation}
\mathcal{Q} = \frac{\bar{m}}{\bar{s}}.
\end{equation}
A region is substructured if $\mathcal{Q} < 0.8$, and centrally concentrated if $\mathcal{Q} > 0.8$. We show the evolution of $\mathcal{Q}$ in our simulations compared to the measured values in Fig.~\ref{Qpar} at various ages (see Table~\ref{summary_regions} for a key to the symbols). The determination of the $\mathcal{Q}$--parameter requires only positional information; however, it can be affected by extinction and membership uncertainty \citep{Bastian09,Parker12d}. Where there is an uncertainty associated with the determination of $\mathcal{Q}$, we show the likely direction of the uncertainty. For example, \citet{Cartwright04} determined $\mathcal{Q} = 0.85$ for $\rho$~Oph; however, using an updated census discussed in \citet{Alves12}, \citet{Parker12c} find $\mathcal{Q} = 0.56$. In our subsequent analysis, we consider any evolutionary scenario that is consistent with either value to be plausible initial conditions for that star-forming region. Similarly, depending on membership probabality, Upper Sco and CrA may have lower $\mathcal{Q}$--parameters (once probable back- and foreground stars are removed), whereas the ONC likely has a higher $\mathcal{Q}$--parameter than that determined from the \citet{Hillenbrand98} data due to visual extinction and sample incompleteness. Finally, Cham~I is slightly elongated, which means the true $\mathcal{Q}$--parameter is slightly higher than measured \citep[0.71 instead of 0.66,][]{Cartwright09a}. These `alternative' measurements are shown in column~4 of Table~\ref{summary_regions} ($\mathcal{Q}_{\rm alt.}$).     

As in Fig.~\ref{density}, panel (a) shows the simulations with initially high densities, panel (b) shows the medium density simulations and panel (c) shows the low density simulations. The solid grey lines are the individual simulations, and the horizontal red dashed line shows the boundary between substructured regions ($\mathcal{Q} < 0.8$) and centrally concentrated regions ($\mathcal{Q} > 0.8$).

\begin{figure*}
  \begin{center}
\setlength{\subfigcapskip}{10pt}
\hspace*{-0.7cm}
\subfigure[High initial density ($\tilde{\rho} \sim10^4$\,M$_\odot$\,pc$^{-3}$)]{\label{Q_Sig-a}\rotatebox{270}{\includegraphics[scale=0.28]{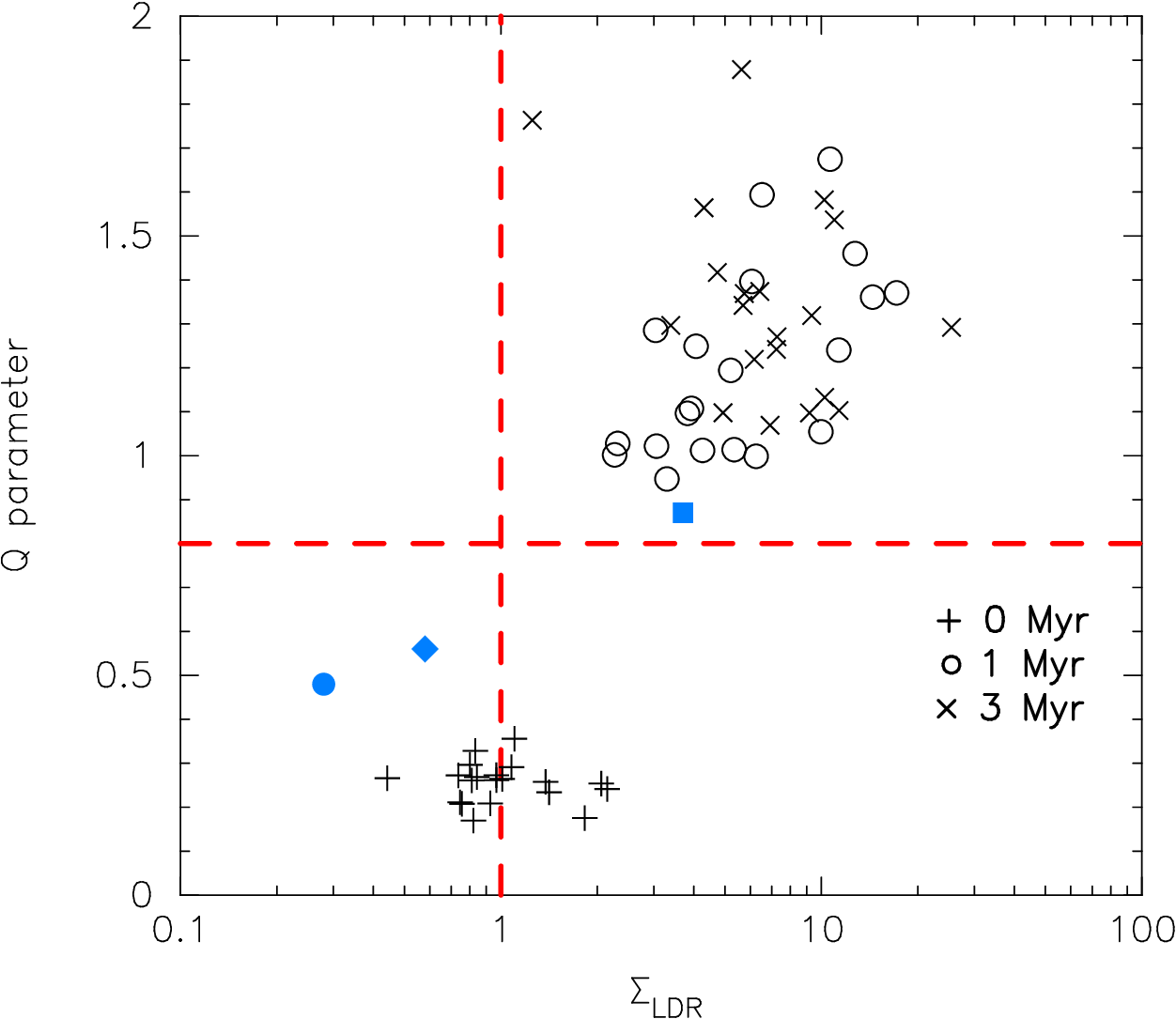}}}  
\hspace*{0.1cm}
\subfigure[Medium initial density ($\tilde{\rho} \sim10^2$\,M$_\odot$\,pc$^{-3}$)]{\label{Q_Sig-b}\rotatebox{270}{\includegraphics[scale=0.28]{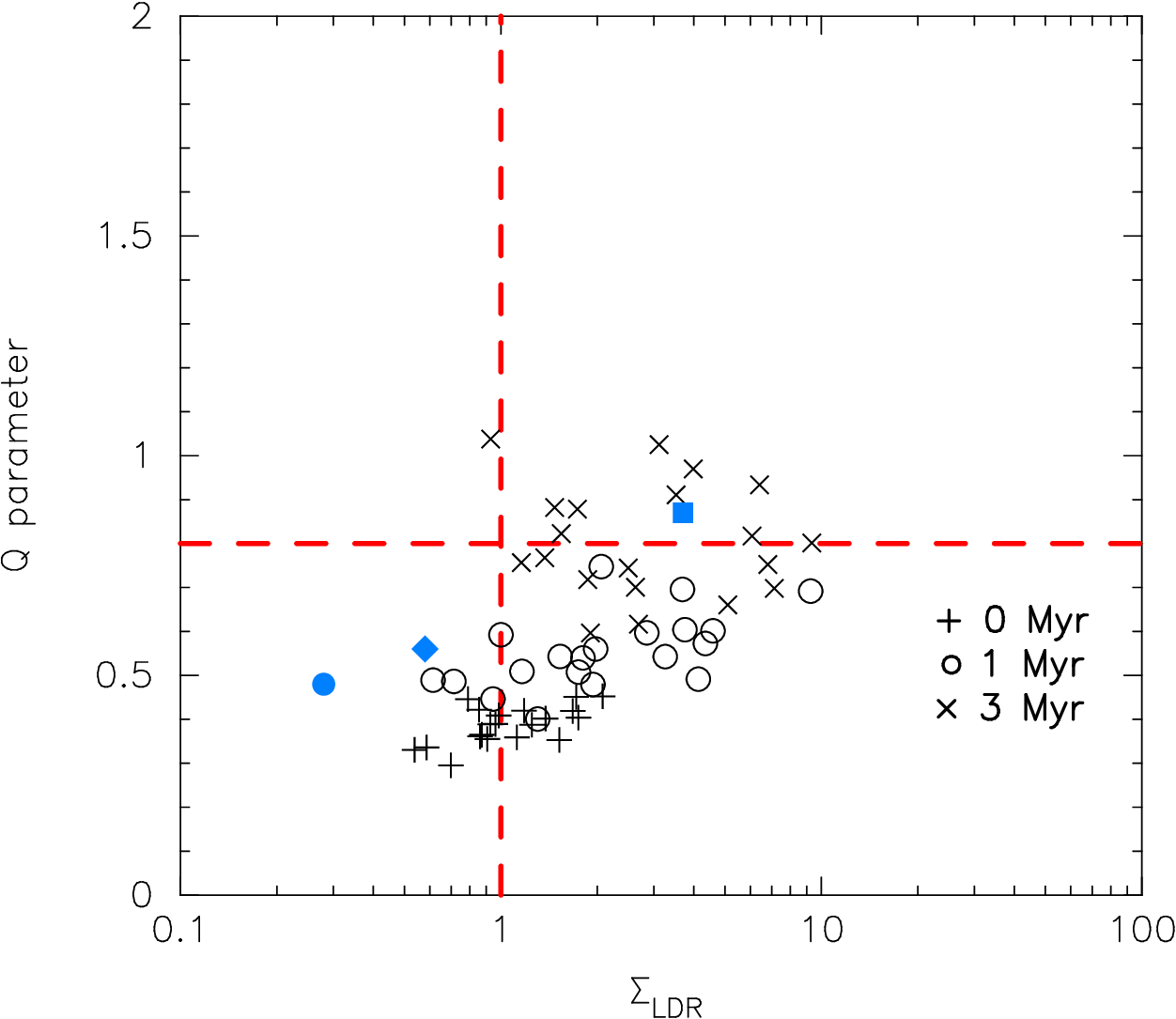}}}
\hspace*{0.1cm}
\subfigure[Low initial density ($\tilde{\rho} \sim10$\,M$_\odot$\,pc$^{-3}$)]{\label{Q_Sig-c}\rotatebox{270}{\includegraphics[scale=0.28]{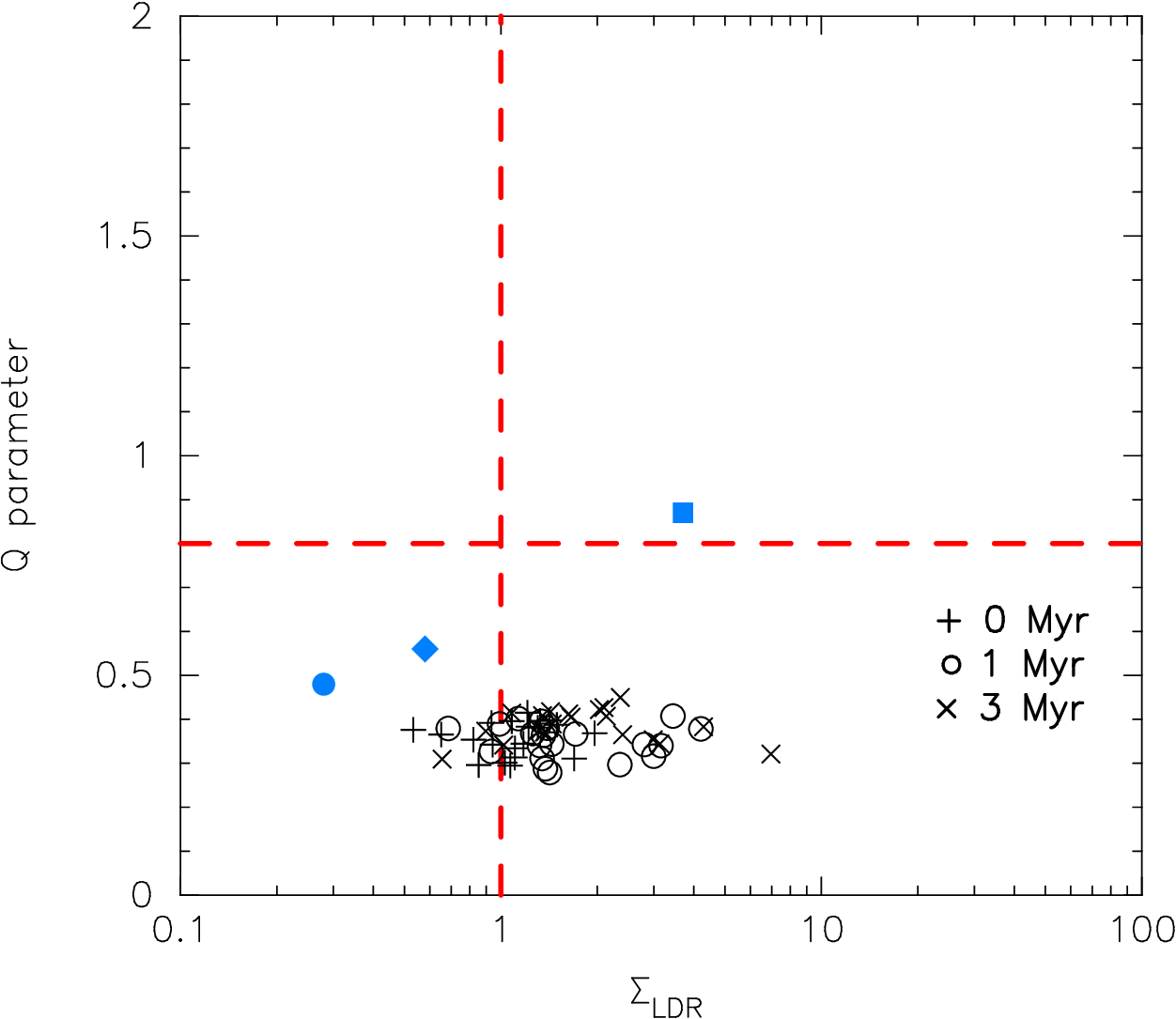}}}
\end{center}
  \caption[bf]{Evolution of structure as measured by the $\mathcal{Q}$-parameter in our simulated star-forming regions versus the relative local density around massive stars compared to the region's median ($\Sigma_{\rm LDR}$). We show values at 0\,Myr (plus signs), 1\,Myr (open circles) and 3\,Myr (crosses). We show the observed values for Taurus (the filled circle), $\rho$~Oph (the filled diamond) and the ONC (the filled square). In panel (a) the star-forming regions have high initial densities ($\tilde{\rho} \sim10^4$M$_\odot$\,pc$^{-3}$), in panel (b) the regions have medium initial densities ($\tilde{\rho} \sim10^2$\,M$_\odot$\,pc$^{-3}$) and in panel (c) the regions have much lower initial densities ($\tilde{\rho} \sim10$\,M$_\odot$\,pc$^{-3}$). The boundary between substructured regions and centrally concentrated regions at  $\mathcal{Q} = 0.8$ is shown by the horizontal dashed line, and $\Sigma_{\rm LDR} = 1$ (where the median local density around massive stars is equal to the region median) is shown by the vertical dashed line.}   
  \label{Q_Sig}
\end{figure*}

We exclude unbound stars from the determination of $\mathcal{Q}$ for two reasons. Firstly,  $\mathcal{Q}$ can appear artifically high when distant stars are included in the analysis, and secondly, stars that are unbound in the simulations are likely to travel far from the regions very quickly, making the comparison with observations unfair. 

As pointed out in \citet{Parker12d} and \citet{Parker14b}, the more dense a region is initially, the more readily substructure is erased, and this is apparent in Fig.~\ref{Qpar}. The most dense regions lose substructure within 1\,Myr (panel a), the medium density regions lose substructure within 5\,Myr (panel b) and the low density regions retain substructure for the duration of the simulations (panel c). Given the high initial densities in Fig.~\ref{Qpar-a}, only the ONC is consistent with very dense initial conditions. When the initial conditions are a factor of $\sim$100 less dense, the measured   $\mathcal{Q}$--parameters for every region apart from the ONC are consistent with more quiescent, medium density initial conditions for star formation. 

\subsection{The $\mathcal{Q}-\Sigma_{\rm LDR}$ plot}

Finally, we present the $\mathcal{Q}-\Sigma_{\rm LDR}$ plot \citep{Parker14b} for our simulations in Fig.~\ref{Q_Sig}. This combines the $\mathcal{Q}$--parameter with the ratio of the median surface density of the 10 most massive stars compared to the median surface density of the region as a whole \citep{Maschberger11}; 
\begin{equation}
\Sigma_{\rm LDR} = \frac{\tilde{\Sigma}_{10}}{\tilde{\Sigma}_{\rm all}}.
\end{equation} 
In Fig.~\ref{Q_Sig} we show the datapoints for Taurus (filled circle), $\rho$~Oph (filled diamond) and the ONC (filled square). These are the only regions in our sample for which we have a reliable census with mass estimates for each individual star in order to determine $\Sigma_{\rm LDR}$. 

Under the reasonable assumption that the velocities of stars are correlated on local scales \citep{Larson82}, \citet{Parker14b} showed that massive stars attain higher surface densities than the median in the region, because they act as potential wells and acquire a retinue of low-mass stars. In the high density simulated regions (Fig.~\ref{Q_Sig-a}), all of the simulations develop high $\Sigma_{\rm LDR}$ values in addition to erasing the primordial substructure. The only observed region which is consistent with these initial conditions is the ONC, and this appears to be marginal. The other observed regions (Taurus and $\rho$~Oph) are more consistent with a much lower initial density, as they have $\Sigma_{\rm LDR} < 1$ and $\mathcal{Q} < 0.8$.

\subsection{Discussion of individual regions}

Recently, \citet{King12b} claimed that differences between the binary separation distributions in nearby star-forming regions were likely to be primordial, as the main differences between binary populations in some regions and the corresponding separation range in the Galactic field were in the `hard' binary regime ($<100$\,au), and thus unlikely to be the result of dynamical evolution. However, \citet{Marks14} show that when the binary fraction is also considered, all of the regions dicussed in \citet{King12b} are in fact consistent with the dynamical evolution of a common binary population and the observed differences between regions are likely due to those regions having different initial densities. Here, we combine the results shown in Figs.~\ref{density},~\ref{Qpar}~and~\ref{Q_Sig} to determine the likely initial (or maximum) density of each region in Table~\ref{summary_regions} \citep[the star forming regions presented in][]{King12b}, and whether this density is consistent with dynamical processing of the universal initial binary population, as suggested by \citet{Marks12} and \citet{Marks14}. We summarise the results in Table~\ref{limits_regions}.

\begin{table*}
\caption[bf]{Comparison of the structure and density of seven star-forming regions with $N$-body simulations to determine which are compatible with the universal initial binary population from \citet{Kroupa95a,Kroupa95b}. From left to right, the columns show the region name, age, $\mathcal{Q}$--parameter (where two values are given due to observational uncertainty, the arrow indicates the more likely value), the observed present-day density of each region as noted by \citet{Marks12} and \citet{King12b}, $\rho_{\rm obs.}$, the postulated initial density with upper and lower limits from \citet{Marks12} and \citet{Marks14} for the binary population of that region to be consistent with the universal primordial binary properties \citep{Kroupa95a,Kroupa95b}, $\rho_{\rm post.}$, the maximum possible initial density when the $\mathcal{Q}$--parameter is also considered, $\rho_{\rm max.}$, and whether or not this region is consistent with the \citet{Kroupa95a,Kroupa95b} universal initial binary population.}
\begin{center}
\begin{tabular}{|c|c|c|c|c|c|c|}
\hline 
Region & Age & $\mathcal{Q}$ & $\rho_{\rm obs.}$ & $\rho_{\rm post.}$ & $\rho_{\rm max.}$ & Universal population? \\
\hline
ONC & 1\,Myr & 0.87 $\rightarrow$ 0.94 & 400\,M$_\odot$\,pc$^{-3}$  & $68\,000^{+90\,000}_{-46\,000}$\,M$_\odot$\,pc$^{-3}$ & 10\,000\,M$_\odot$\,pc$^{-3}$ & no$^a$ \vspace*{0.15cm}\\
$\rho$~Oph & 1\,Myr & 0.56 $\leftarrow$ 0.85 & 200\,M$_\odot$\,pc$^{-3}$ & $2300^{+7400}_{-1100}$\,M$_\odot$\,pc$^{-3}$ & 2000\,M$_\odot$\,pc$^{-3}$ & yes \vspace*{0.15cm}\\  
Taurus & 1\,Myr & 0.48 & 8\,M$_\odot$\,pc$^{-3}$  & $350^{+850}_{-140}$\,M$_\odot$\,pc$^{-3}$ &  10\,M$_\odot$\,pc$^{-3}$ & no \vspace*{0.15cm}\\
IC\,348 & 3\,Myr & 0.92 & 180\,M$_\odot$\,pc$^{-3}$ & $9400^{+53\,000}_{-2700}$\,M$_\odot$\,pc$^{-3}$ &  1000\,M$_\odot$\,pc$^{-3}$ & no \vspace*{0.15cm}\\ 
Cham~I & 3\,Myr & 0.66 $\rightarrow$ 0.71 & 1\,M$_\odot$\,pc$^{-3}$ & $1600^{+13\,000}_{-230}$\,M$_\odot$\,pc$^{-3}$ & 1\,M$_\odot$\,pc$^{-3}$& no \vspace*{0.15cm} \\
CrA & 1\,Myr & 0.32 $\leftarrow$ 0.38 & 30\,M$_\odot$\,pc$^{-3}$ & 190\,M$_\odot$\,pc$^{-3}$ & 100\,M$_\odot$\,pc$^{-3}$ & yes \vspace*{0.15cm}  \\
Upper Sco & 5\,Myr & 0.75 $\leftarrow$ 0.88 &  16\,M$_\odot$\,pc$^{-3}$ & 4200\,M$_\odot$\,pc$^{-3}$ & 1000\,M$_\odot$\,pc$^{-3}$ & no \\
\hline
\end{tabular}
\end{center}
\begin{flushleft}
 $^a$Note that an initial density of $10^4$\,M$_\odot$\,pc$^{-3}$ for the ONC appears to be consistent with the universal binary population in Fig.~\ref{separations}. However, this (relatively low) density was ruled out at 90\,per cent confidence by \citet{Marks12}.
\end{flushleft}
\label{limits_regions}
\end{table*}

{\bf ONC:} The ONC has both a high $\mathcal{Q}$--parameter and $\Sigma_{\rm LDR}$ ratio, which suggests that its initial density was likely $\tilde{\rho} \sim10^4$\,M$_\odot$\,pc$^{-3}$. However, if its density were higher, the $\mathcal{Q}$--parameter would also be higher, so $10^4$\,M$_\odot$\,pc$^{-3}$ is very much an upper limit on the initial density. \citet{Marks12} suggest an initial density of 68\,000\,M$_\odot$\,pc$^{-3}$, with values lower than  46\,000\,M$_\odot$\,pc$^{-3}$ or higher than  90\,000\,M$_\odot$\,pc$^{-3}$ excluded with 90\,per cent confidence. However, we note that the evolution of the universal binary population in our regions does appear to be consistent with the data from \citet{Reipurth07} -- compare the solid histogram with the squares in Fig.~\ref{separations}.  

{\bf $\rho$~Oph:} If we take  $\mathcal{Q} = 0.56$ \citep{Alves12,Parker12c} for $\rho$~Oph, with $\Sigma_{\rm LDR} = 0.58$ then this is consistent with a moderate initial density ($10^2 - 10^3$\,M$_\odot$\,pc$^{-3}$), rather than high initial (or maximum densities). \citet{Marks12} suggest an initial density of 2300\,M$_\odot$\,pc$^{-3}$, with values lower than  1100\,M$_\odot$\,pc$^{-3}$ or higher than  7400\,M$_\odot$\,pc$^{-3}$ excluded with 90\,per cent confidence. One of our medium density simulations briefly reaches a density of 2000\,M$_\odot$\,pc$^{-3}$, suggesting that the evolution of this region (and processing of binaries) could be consistent with the universal initial binary population.

{\bf Taurus:} The low $\mathcal{Q}$--parameter \citep[0.48 --][]{Cartwright04} and low $\Sigma_{\rm LDR}$ \citep[0.28 using the dataset from][]{Parker11b} suggest that dynamical evolution has not been significant in this region. Taurus is consistent with very quiescent initial conditions ($\rho \sim 10$\,M$_\odot$\,pc$^{-3}$ -- Figs.~\ref{density-c}~and~\ref{Qpar-c}).  \citet{Marks12} suggest an initial density of 350\,M$_\odot$\,pc$^{-3}$, with values lower than  140\,M$_\odot$\,pc$^{-3}$ or higher than  850\,M$_\odot$\,pc$^{-3}$ excluded with 90\,per cent confidence. Given its current low density, low $\mathcal{Q}$--parameter and low $\Sigma_{\rm LDR}$, Taurus is not consistent with the universal initial binary population. 

{\bf IC\,348:} Because of its age (3\,Myr), if IC\,348 had initially high density, two-body relaxation would have reduced the density to values much lower than observed for this region (see Fig.~\ref{density}). This, combined with the $\mathcal{Q}$--parameter of 0.92 \citep{Cartwright04} suggests a moderate initial density ($10^2 - 10^3$\,M$_\odot$\,pc$^{-3}$ -- see panel (b) of Figs.~\ref{density}~and~\ref{Qpar}). \citet{Marks12} suggest an initial density of 9400\,M$_\odot$\,pc$^{-3}$, with values lower than  2700\,M$_\odot$\,pc$^{-3}$ or higher than  53\,000\,M$_\odot$\,pc$^{-3}$ excluded with 90\,per cent confidence. Such a high initial density is inconsistent with the observed structure and current density, and its binary population is probably not evolved from the \citet{Kroupa95b} universal binary population. 

{\bf Cham~I:} Chamaeleon~I has an age of 3\,Myr, and a low density, but relatively high  $\mathcal{Q}$--parameter of 0.71 \citep{Cartwright09a}. Figs.~\ref{density}~and~\ref{Qpar} show that none of our dynamical scenarios fit the observed values, and it is therefore likely that Cham~I formed with its current density and structure -- and that dynamical evolution has not altered its binary population.   \citet{Marks12} suggest an initial density of 1600\,M$_\odot$\,pc$^{-3}$, with values lower than  230\,M$_\odot$\,pc$^{-3}$ or higher than  13\,000\,M$_\odot$\,pc$^{-3}$ excluded with 90\,per cent confidence. Given its current low density and lack of dynamical evolution, Cham~I is not consistent with the universal binary population. 

{\bf CrA:} CrA has $\mathcal{Q} = 0.38$ but a moderate density of 30\,M$_\odot$\,pc$^{-3}$ \citep{Neuhauser08,King12b}. According to our evolutionary models, CrA could have evolved slightly from a highly substructured region with an initial density of $\sim~10^2$\,M$_\odot$\,pc$^{-3}$. In order to be consistent with the initial universal binary population, \citet{Marks14} suggest an initial density of 190\,M$_\odot$\,pc$^{-3}$ (without providing limits). Our analysis suggests that this region is consistent with the universal initial binary population, assuming a similar magnitude in the confidence limit range as for the other regions. 

{\bf Upper Sco:} The $\mathcal{Q}$--parameter is 0.88 \citep{Kraus07b} and current density is 16\,M$_\odot$\,pc$^{-3}$ \citep{King12b}; both of which imply that Upper Sco is likely to have had moderately dense initial conditions  ($\rho \sim 10^2 - 10^3$\,M$_\odot$\,pc$^{-3}$). \citet{Marks14} suggest an initial density of 4200\,M$_\odot$\,pc$^{-3}$ (again without confidence limits). Our evolutionary models suggest that Upper Sco is also inconsistent with the initial densities required to process the universal initial binary population.  \\

In summary, only $\rho$~Oph, CrA, and possibly the ONC, are consistent with the dynamical processing of the universal initial binary population from \citet{Kroupa95a,Kroupa95b}, based on the consideration of the regions' structure and current density. We suggest that the observed differences between these regions are likely to be a relic of the star formation process, although we caution that as the binaries observed in these regions are `intermediate' they may still have undergone some degree of dynamical evolution \citep{Parker12b}. 
 
\section{Conclusions}
\label{conclude}

 We have presented $N$-body simulations of the dynamical evolution of star-forming regions in which we follow the stellar density and spatial structure and compare the results to seven observed regions. For each individual region, we determine the likely initial density (which is usually, but not always, the maximum) based on its observed current density and spatial structure, as determined by the $\mathcal{Q}$--parameter.

The spatial structure of a region is a strong constraint on the amount of dynamical evolution that has taken place, as dense regions $\tilde{\rho} > 10^3$\,M$_\odot$\,pc$^{-3}$ erase structure almost immediately, intermediate density regions ($\tilde{\rho} \sim 10^2 - 10^3$\,M$_\odot$\,pc$^{-3}$) remove structure within 5\,Myr but low-density regions ($\tilde{\rho} < $\,10\,M$_\odot$\,pc$^{-3}$) retain structure beyond the age of all of the regions considered here. Folding in the measurement of structure largely removes the density degeneracy problem in star formation, where the initial density is very difficult to constrain due to the rapid expansion of initially dense regions, and the slower expansion of more quiescent regions, both of which can result in the same present-day density from very different initial conditions.  

Our results can be used to infer the likely maximum density of observed star-forming regions, which for example enables the importance of the effects of dynamical interactions and radiation from massive stars on protoplanetary discs to be ascertained \citep[e.g.][]{Scally01,Adams06,Rosotti14}. Recently, \citet{deJuan12} showed an apparent dependence of the size of protoplanetary discs on the density of the star-forming environment, although their observations were limited to nearby star-forming regions. Future ALMA observations may be able to probe discs in more distant regions \citep[e.g.][]{Mann14}, and using the $\mathcal{Q}$--parameter in tandem with the present-day density will be useful in determining whether any observed trends in disc size are due to the star-formation environment. 

We also apply our method to determine which of seven nearby star-forming regions are consistent with the `universal initial binary population' model for star formation \citep{Kroupa95a,Kroupa95b}, based on recent numerical simulations presented in the literature \citep{Marks12,Marks14}. We compare the density of our simulations which fit the observed regions' structure and determine whether the initial density of those simulations is high enough to process the initial binary population to resemble the binary properties observed in each region today, using the values quoted in \citet{Marks12} and \citet{Marks14}. We find that of the seven regions observed, only three -- $\rho$~Oph, CrA and possibly the ONC -- are consistent with the universal initial binary population model for star formation. 

Unfortunately, aside from discarding the universal initial binary population hypothesis in \citet{Kroupa95a,Kroupa95b}, our results do not help much in assessing the type of star-forming region that contributes binaries to the Galactic field. We are still limited by the observed separation range in regions (10s -- 1000s\,au) which is small compared to the field ($10^{-2} - 10^5$\,au), and by the fact that these visual binaries are dynamically `intermediate' systems \citep{Heggie75,Hills75a,Hills75b} that could have evolved stochastically \citep[especially in dense regions like the ONC,][]{Parker12b}. We also have very little information on whether the regions considered are representative of those that do populate the field. Further observations of e.g.\,\,spectroscopic binaries which are not affected by dynamical evolution are desperately required in order to look for stark differences between the binary populations of the regions in question and the Galactic field.

\section*{Acknowledgements}

I am grateful to the referee, Cathie Clarke, for her comments and suggestions which have led to a more interesting paper. I acknowledge support from the Royal Astronomical Society in the form of a research fellowship.

\bibliographystyle{mn2e}
\bibliography{general_ref}

\begin{thebibliography}{}

\bibitem[\protect\citeauthoryear{Adams}{Adams}{2011}]{Adams11}
Adams F.~C.,  2011, ARA\&A, 48, 47

\bibitem[\protect\citeauthoryear{Adams, Proszkow, Fatuzzo \& Myers}{Adams
  et~al.}{2006}]{Adams06}
Adams F.~C.,  Proszkow E.~M.,  Fatuzzo M.,    Myers P.~C.,  2006, ApJ, 641, 504

\bibitem[\protect\citeauthoryear{{Alexander}, {Pascucci}, {Andrews}, {Armitage}
  \& {Cieza}}{{Alexander} et~al.}{2013}]{Alexander13}
{Alexander} R.,  {Pascucci} I.,  {Andrews} S.,  {Armitage} P.,    {Cieza} L.,
  2013, arXiv: 1311.1819

\bibitem[\protect\citeauthoryear{Allison, Goodwin, Parker, {Portegies Zwart} \&
  de Grijs}{Allison et~al.}{2010}]{Allison10}
Allison R.~J.,  Goodwin S.~P.,  Parker R.~J.,  {Portegies Zwart} S.~F.,    de
  Grijs R.,  2010, MNRAS, 407, 1098

\bibitem[\protect\citeauthoryear{{Alves de Oliveira}, {Moraux}, {Bouvier} \&
  {Bouy}}{{Alves de Oliveira} et~al.}{2012}]{Alves12}
{Alves de Oliveira} C.,  {Moraux} E.,  {Bouvier} J.,    {Bouy} H.,  2012, A\&A,
  539, A151

\bibitem[\protect\citeauthoryear{{Andr{\'e}}, {Men'shchikov}, {Bontemps},
  {K{\"o}nyves}, {Motte}, {Schneider}, {Didelon}, {Minier}, {Saraceno},
  {Ward-Thompson} \& {et al.}}{{Andr{\'e}} et~al.}{2010}]{Andre10}
{Andr{\'e}} P.,  {Men'shchikov} A.,  {Bontemps} S.,  {K{\"o}nyves} V.,  {Motte}
  F.,  {Schneider} N.,  {Didelon} P.,  {Minier} V.,  {Saraceno} P.,
  {Ward-Thompson} D.,    {et al.} 2010, A\&A, 518, L102

\bibitem[\protect\citeauthoryear{{Armitage}}{{Armitage}}{2000}]{Armitage00}
{Armitage} P.~J.,  2000, A\&A, 362, 968

\bibitem[\protect\citeauthoryear{Bastian, Gieles, Ercolano \&
  Gutermuth}{Bastian et~al.}{2009}]{Bastian09}
Bastian N.,  Gieles M.,  Ercolano B.,    Gutermuth R.,  2009, MNRAS, 392, 868

\bibitem[\protect\citeauthoryear{Bonnell, Bate, Clarke \& Pringle}{Bonnell
  et~al.}{2001}]{Bonnell01}
Bonnell I.~A.,  Bate M.~R.,  Clarke C.~J.,    Pringle J.~E.,  2001, MNRAS, 323,
  785

\bibitem[\protect\citeauthoryear{Bressert, Bastian, Gutermuth, Megeath, Allen,
  {Evans, II}, Rebull, Hatchell, Johnstone, Bourke, Cieza, Harvey, Merin, Ray
  \& Tothill}{Bressert et~al.}{2010}]{Bressert10}
Bressert E.,  Bastian N.,  Gutermuth R.,  Megeath S.~T.,  Allen L.,  {Evans,
  II} N.~J.,  Rebull L.~M.,  Hatchell J.,  Johnstone D.,  Bourke T.~L.,  Cieza
  L.~A.,  Harvey P.~M.,  Merin B.,  Ray T.~P.,    Tothill N. F.~H.,  2010,
  MNRAS, 409, L54

\bibitem[\protect\citeauthoryear{Cartwright}{Cartwright}{2009}]{Cartwright09b}
Cartwright A.,  2009, MNRAS, 400, 1427

\bibitem[\protect\citeauthoryear{Cartwright \& Whitworth}{Cartwright \&
  Whitworth}{2004}]{Cartwright04}
Cartwright A.,  Whitworth A.~P.,  2004, MNRAS, 348, 589

\bibitem[\protect\citeauthoryear{{Cartwright} \& {Whitworth}}{{Cartwright} \&
  {Whitworth}}{2009}]{Cartwright09a}
{Cartwright} A.,  {Whitworth} A.~P.,  2009, MNRAS, 392, 341

\bibitem[\protect\citeauthoryear{{Dale}, {Ercolano} \& {Bonnell}}{{Dale}
  et~al.}{2013}]{Dale13}
{Dale} J.~E.,  {Ercolano} B.,    {Bonnell} I.~A.,  2013, MNRAS, 430, 234

\bibitem[\protect\citeauthoryear{{Davies}, {Adams}, {Armitage}, {Chambers},
  {Ford}, {Morbidelli}, {Raymond} \& {Veras}}{{Davies} et~al.}{2013}]{Davies13}
{Davies} M.~B.,  {Adams} F.~C.,  {Armitage} P.,  {Chambers} J.,  {Ford} E.,
  {Morbidelli} A.,  {Raymond} S.~N.,    {Veras} D.,  2013, arXiv: 1311.6816

\bibitem[\protect\citeauthoryear{{de Juan Ovelar}, Kruijssen, Bressert, Testi,
  Bastian \& {C{\'a}novas Cabrera}}{{de Juan Ovelar} et~al.}{2012}]{deJuan12}
{de Juan Ovelar} M.,  Kruijssen J.,  Bressert E.,  Testi L.,  Bastian N.,
  {C{\'a}novas Cabrera} H.,  2012, A\&A, 546, L1

\bibitem[\protect\citeauthoryear{{Duch{\^e}ne} \& {Kraus}}{{Duch{\^e}ne} \&
  {Kraus}}{2013}]{Duchene13b}
{Duch{\^e}ne} G.,  {Kraus} A.,  2013, ARA\&A, 51, 269

\bibitem[\protect\citeauthoryear{Duquennoy \& Mayor}{Duquennoy \&
  Mayor}{1991}]{Duquennoy91}
Duquennoy A.,  Mayor M.,  1991, A\&A, 248, 485

\bibitem[\protect\citeauthoryear{Fischer \& Marcy}{Fischer \&
  Marcy}{1992}]{Fischer92}
Fischer D.~A.,  Marcy G.~W.,  1992, ApJ, 396, 178

\bibitem[\protect\citeauthoryear{Gieles, Moeckel \& Clarke}{Gieles
  et~al.}{2012}]{Gieles12}
Gieles M.,  Moeckel N.,    Clarke C.~J.,  2012, MNRAS, 426, L11

\bibitem[\protect\citeauthoryear{{Girichidis}, {Federrath}, {Allison},
  {Banerjee} \& {Klessen}}{{Girichidis} et~al.}{2012}]{Girichidis12}
{Girichidis} P.,  {Federrath} C.,  {Allison} R.,  {Banerjee} R.,    {Klessen}
  R.~S.,  2012, MNRAS, 420, 3264

\bibitem[\protect\citeauthoryear{Goodwin}{Goodwin}{2010}]{Goodwin10}
Goodwin S.~P.,  2010, {Royal Society of London Philosophical Transactions
  Series A}, 368, 851

\bibitem[\protect\citeauthoryear{Goodwin \& Whitworth}{Goodwin \&
  Whitworth}{2004}]{Goodwin04a}
Goodwin S.~P.,  Whitworth A.~P.,  2004, A\&A, 413, 929

\bibitem[\protect\citeauthoryear{{Gouliermis}, {Hony} \&
  {Klessen}}{{Gouliermis} et~al.}{2014}]{Gouliermis14}
{Gouliermis} D.~A.,  {Hony} S.,    {Klessen} R.~S.,  2014, MNRAS, 439, 3775

\bibitem[\protect\citeauthoryear{Heggie}{Heggie}{1975}]{Heggie75}
Heggie D.~C.,  1975, MNRAS, 173, 729

\bibitem[\protect\citeauthoryear{Hillenbrand \& Hartmann}{Hillenbrand \&
  Hartmann}{1998}]{Hillenbrand98}
Hillenbrand L.~A.,  Hartmann L.~W.,  1998, ApJ, 492, 540

\bibitem[\protect\citeauthoryear{Hills}{Hills}{1975a}]{Hills75a}
Hills J.~G.,  1975a, AJ, 80, 809

\bibitem[\protect\citeauthoryear{Hills}{Hills}{1975b}]{Hills75b}
Hills J.~G.,  1975b, AJ, 80, 1075

\bibitem[\protect\citeauthoryear{King, Parker, Patience \& Goodwin}{King
  et~al.}{2012a}]{King12a}
King R.~R.,  Parker R.~J.,  Patience J.,    Goodwin S.~P.,  2012a, MNRAS, 421,
  2025

\bibitem[\protect\citeauthoryear{King, Goodwin, Parker \& Patience}{King
  et~al.}{2012b}]{King12b}
King R.~R.,  Goodwin S.~P.,  Parker R.~J.,    Patience J.,  2012b, MNRAS, 427,
  2636

\bibitem[\protect\citeauthoryear{K{\"o}hler \& Leinert}{K{\"o}hler \&
  Leinert}{1998}]{Kohler98}
K{\"o}hler R.,  Leinert C.,  1998, A\&A, 331, 977

\bibitem[\protect\citeauthoryear{{Kraus} \& {Hillenbrand}}{{Kraus} \&
  {Hillenbrand}}{2007}]{Kraus07b}
{Kraus} A.~L.,  {Hillenbrand} L.~A.,  2007, ApJ, 662, 413

\bibitem[\protect\citeauthoryear{Kroupa}{Kroupa}{1995a}]{Kroupa95a}
Kroupa P.,  1995a, MNRAS, 277, 1491

\bibitem[\protect\citeauthoryear{Kroupa}{Kroupa}{1995b}]{Kroupa95b}
Kroupa P.,  1995b, MNRAS, 277, 1507

\bibitem[\protect\citeauthoryear{Kroupa}{Kroupa}{1995c}]{Kroupa95c}
Kroupa P.,  1995c, MNRAS, 277, 1522

\bibitem[\protect\citeauthoryear{Kroupa}{Kroupa}{2002}]{Kroupa02}
Kroupa P.,  2002, Science, 295, 82

\bibitem[\protect\citeauthoryear{Kroupa, Petr \& McCaughrean}{Kroupa
  et~al.}{1999}]{Kroupa99}
Kroupa P.,  Petr M.~G.,    McCaughrean M.~J.,  1999, New Astronomy, 4, 495

\bibitem[\protect\citeauthoryear{Kroupa \& {Petr-Gotzens}}{Kroupa \&
  {Petr-Gotzens}}{2011}]{Kroupa11}
Kroupa P.,  {Petr-Gotzens} M.~G.,  2011, A\&A, 529, A92

\bibitem[\protect\citeauthoryear{Lada}{Lada}{2010}]{Lada10}
Lada C.~J.,  2010, {Royal Society of London Philosophical Transactions Series
  A}, 368, 713

\bibitem[\protect\citeauthoryear{Lada \& Lada}{Lada \& Lada}{2003}]{Lada03}
Lada C.~J.,  Lada E.~A.,  2003, ARA\&A, 41, 57

\bibitem[\protect\citeauthoryear{Larson}{Larson}{1982}]{Larson82}
Larson R.~B.,  1982, MNRAS, 200, 159

\bibitem[\protect\citeauthoryear{Leinert, Zinnecker, Weitzel, Christou,
  Ridgway, Jameson, Haas \& Lenzen}{Leinert et~al.}{1993}]{Leinert93}
Leinert C.,  Zinnecker H.,  Weitzel N.,  Christou J.,  Ridgway S.~T.,  Jameson
  R.,  Haas M.,    Lenzen R.,  1993, A\&A, 278, 129

\bibitem[\protect\citeauthoryear{{Mann}, {Di Francesco}, {Johnstone},
  {Andrews}, {Williams}, {Bally}, {Ricci}, {Hughes} \& {Matthews}}{{Mann}
  et~al.}{2014}]{Mann14}
{Mann} R.~K.,  {Di Francesco} J.,  {Johnstone} D.,  {Andrews} S.~M.,
  {Williams} J.~P.,  {Bally} J.,  {Ricci} L.,  {Hughes} A.~M.,    {Matthews}
  B.~C.,  2014, ApJ, 784, 82

\bibitem[\protect\citeauthoryear{{Marks} \& {Kroupa}}{{Marks} \&
  {Kroupa}}{2012}]{Marks12}
{Marks} M.,  {Kroupa} P.,  2012, A\&A, 543, A8

\bibitem[\protect\citeauthoryear{{Marks}, {Leigh}, {Giersz}, {Pfalzner},
  {Pflamm-Altenburg} \& {Oh}}{{Marks} et~al.}{2014}]{Marks14}
{Marks} M.,  {Leigh} N.,  {Giersz} M.,  {Pfalzner} S.,  {Pflamm-Altenburg} J.,
    {Oh} S.,  2014, MNRAS, 441, 3503

\bibitem[\protect\citeauthoryear{Maschberger \& Clarke}{Maschberger \&
  Clarke}{2011}]{Maschberger11}
Maschberger T.,  Clarke C.~J.,  2011, MNRAS, 416, 541

\bibitem[\protect\citeauthoryear{Mathieu}{Mathieu}{1994}]{Mathieu94}
Mathieu R.~D.,  1994, ARA\&A, 32, 465

\bibitem[\protect\citeauthoryear{Metchev \& Hillenbrand}{Metchev \&
  Hillenbrand}{2009}]{Metchev09}
Metchev S.~A.,  Hillenbrand L.~A.,  2009, ApJS, 181, 62

\bibitem[\protect\citeauthoryear{Moeckel, Holland, Clarke \& Bonnell}{Moeckel
  et~al.}{2012}]{Moeckel12}
Moeckel N.,  Holland C.,  Clarke C.~J.,    Bonnell I.~A.,  2012, MNRAS, 425,
  450

\bibitem[\protect\citeauthoryear{{Neuh{\"a}user} \& {Forbrich}}{{Neuh{\"a}user}
  \& {Forbrich}}{2008}]{Neuhauser08}
{Neuh{\"a}user} R.,  {Forbrich} J.,  2008, {The Corona Australis Star Forming
  Region}.
p.~735

\bibitem[\protect\citeauthoryear{Olczak, Pfalzner \& Eckart}{Olczak
  et~al.}{2008}]{Olczak08}
Olczak C.,  Pfalzner S.,    Eckart A.,  2008, A\&A, 488, 191

\bibitem[\protect\citeauthoryear{Parker, Bouvier, Goodwin, Moraux, Allison,
  Guieu \& G{\"u}del}{Parker et~al.}{2011}]{Parker11b}
Parker R.~J.,  Bouvier J.,  Goodwin S.~P.,  Moraux E.,  Allison R.~J.,  Guieu
  S.,    G{\"u}del M.,  2011, MNRAS, 412, 2489

\bibitem[\protect\citeauthoryear{Parker \& Goodwin}{Parker \&
  Goodwin}{2012}]{Parker12b}
Parker R.~J.,  Goodwin S.~P.,  2012, MNRAS, 424, 272

\bibitem[\protect\citeauthoryear{Parker, Maschberger \& {Alves de
  Oliveira}}{Parker et~al.}{2012}]{Parker12c}
Parker R.~J.,  Maschberger T.,    {Alves de Oliveira} C.,  2012, MNRAS, 426,
  3079

\bibitem[\protect\citeauthoryear{Parker \& Meyer}{Parker \&
  Meyer}{2012}]{Parker12d}
Parker R.~J.,  Meyer M.~R.,  2012, MNRAS, 427, 637

\bibitem[\protect\citeauthoryear{Parker \& Quanz}{Parker \&
  Quanz}{2012}]{Parker12a}
Parker R.~J.,  Quanz S.~P.,  2012, MNRAS, 419, 2448

\bibitem[\protect\citeauthoryear{Parker, Wright, Goodwin \& Meyer}{Parker
  et~al.}{2014}]{Parker14b}
Parker R.~J.,  Wright N.~J.,  Goodwin S.~P.,    Meyer M.~R.,  2014, MNRAS, 438,
  620

\bibitem[\protect\citeauthoryear{{Pecaut}, {Mamajek} \& {Bubar}}{{Pecaut}
  et~al.}{2012}]{Pecaut12}
{Pecaut} M.~J.,  {Mamajek} E.~E.,    {Bubar} E.~J.,  2012, ApJ, 746, 154

\bibitem[\protect\citeauthoryear{{Portegies Zwart}, McMillan, Hut \&
  Makino}{{Portegies Zwart} et~al.}{2001}]{Zwart01}
{Portegies Zwart} S.~F.,  McMillan S. L.~W.,  Hut P.,    Makino J.,  2001,
  MNRAS, 321, 199

\bibitem[\protect\citeauthoryear{{Portegies Zwart}, Makino, McMillan \&
  Hut}{{Portegies Zwart} et~al.}{1999}]{Zwart99}
{Portegies Zwart} S.~F.,  Makino J.,  McMillan S. L.~W.,    Hut P.,  1999,
  A\&A, 348, 117

\bibitem[\protect\citeauthoryear{Raghavan, McMaster, Henry, Latham, Marcy,
  Mason, Gies, White \& {ten Brummelaar}}{Raghavan et~al.}{2010}]{Raghavan10}
Raghavan D.,  McMaster H.~A.,  Henry T.~J.,  Latham D.~W.,  Marcy G.~W.,  Mason
  B.~D.,  Gies D.~R.,  White R.~J.,    {ten Brummelaar} T.~A.,  2010, ApJSS,
  190, 1

\bibitem[\protect\citeauthoryear{{Reggiani} \& {Meyer}}{{Reggiani} \&
  {Meyer}}{2013}]{Reggiani13}
{Reggiani} M.~M.,  {Meyer} M.~R.,  2013, A\&A, 553, A124

\bibitem[\protect\citeauthoryear{Reipurth, Guimar{\~a}es, Connelley \&
  Bally}{Reipurth et~al.}{2007}]{Reipurth07}
Reipurth B.,  Guimar{\~a}es M.~M.,  Connelley M.~S.,    Bally J.,  2007, AJ,
  134, 2272

\bibitem[\protect\citeauthoryear{{Rosotti}, {Dale}, {de Juan Ovelar}, {Hubber},
  {Kruijssen}, {Ercolano} \& {Walch}}{{Rosotti} et~al.}{2014}]{Rosotti14}
{Rosotti} G.~P.,  {Dale} J.~E.,  {de Juan Ovelar} M.,  {Hubber} D.~A.,
  {Kruijssen} J.~M.~D.,  {Ercolano} B.,    {Walch} S.,  2014, MNRAS, 441, 2094

\bibitem[\protect\citeauthoryear{S{\'a}nchez \& Alfaro}{S{\'a}nchez \&
  Alfaro}{2009}]{Sanchez09}
S{\'a}nchez N.,  Alfaro E.~J.,  2009, ApJ, 696, 2086

\bibitem[\protect\citeauthoryear{Scally \& Clarke}{Scally \&
  Clarke}{2001}]{Scally01}
Scally A.,  Clarke C.,  2001, MNRAS, 325, 449

\bibitem[\protect\citeauthoryear{{Schmeja} \& {Klessen}}{{Schmeja} \&
  {Klessen}}{2006}]{Schmeja06}
{Schmeja} S.,  {Klessen} R.~S.,  2006, A\&A, 449, 151

\bibitem[\protect\citeauthoryear{Shu, Adams \& Lizano}{Shu
  et~al.}{1987}]{Shu87}
Shu F.~H.,  Adams F.~C.,    Lizano S.,  1987, ARA\&A, 25, 23

\end{thebibliography}

\label{lastpage}

\end{document}